\documentclass{iau} 
\usepackage{graphicx}

\title[~~ Astro-WISE - Target Data Federations ] %% give here short title 
{Target and  (Astro-)WISE technologies\\
Data federations and its applications}

\author[Edwin A. Valentijn, \etal]{E.A. Valentijn$^1$, 
K. Begeman$^1$, A. Belikov$^1$, D.R. Boxhoorn$^1$,\\ 
J. Brinchmann$^2$, J. McFarland$^1$, H. Holties$^3$, K.H. Kuijken$^2$,\\
G. Verdoes Kleijn$^1$, W-J. Vriend$^1$, O.R. Williams$^4$,\\ J.B.T.M. Roerdink$^5$, L.R.B. Schomaker$^6$, M.A. Swertz$^7$,\\
A.Tsyganov$^4$, G.J.W. van Dijk$^8$}

\affiliation{$^1$ Kapteyn Astronomical Institute, University of Groningen, \\ 
email: {\tt valentyn@astro.rug.nl}\\$^2$ Leiden Observatory, Leiden University \\$^3$ ASTRON, Dwingeloo\\$^4$ Center for Information Technology, University of Groningen \\ 
$^5$ Johann Bernoulli Institute, University of Groningen \\$^6$ ALICE, University of Groningen \\
$^7$ University Medical Center Groningen, University of Groningen \\$^8$ Target Holding, Groningen}
\pubyear{2017}
\volume{325}  %% insert here IAU Symposium No.
\jname{Astroinformatics}
\editors{M.C Brescia,  \& G. Longo, eds.}

\begin{document}

\maketitle

\begin{abstract}
After its first implementation in 2003 the Astro-WISE technology has been rolled out in several European countries  and is used for the production of the KiDS survey data. In the multi-disciplinary Target initiative this technology, nicknamed WISE technology, has been further applied to a large number of projects. Here, we highlight the data handling of other astronomical applications, such as VLT-MUSE and LOFAR,  together with some non-astronomical applications such as the medical projects Lifelines and GLIMPS; the MONK handwritten text recognition system; and business applications, by amongst others, the Target Holding.\\
We describe some of the most important lessons learned and describe the application of the data-centric  WISE type of approach to the Science Ground Segment of the Euclid satellite.
\end{abstract}

\firstsection % if your document starts with a section,
\section{Introduction- the data-centric and distributed approach}
The Astro-WISE information system ({http://www.astro-wise.org), the first distributed and data-centric end-to-end information system in astronomy, saw its first light in 2003.  It was originally developed to process, calibrate and quality control  the data of the Kilo  Degree Survey an optical astronomy wide field survey at the VST telescope.  Astro-WISE is a data production system, which has the additional feature that it can publicly publish results, including dependencies. The system builds on a unique approach to the architecture of scientific information systems. The modeling of the data handling is done in a data-centric mode, in which  all data beyond pixel data is identified as Metadata and maintained and distributed by a database, while the true pixel data (the files) are distributed by a data server, who queries the database about the files whereabouts. The actual processing is handled by a distributed processing unit, while at the computer nodes metadata are resolved through database access, while files are resolved by the data servers. This set-up allows to connect various datacenters across Europe, together with their associated satellites, to work simultaneously on a given project; synchronization is maintained in real-time and the whole community has access to all results and all the data, both files and metadata, that went into the product. 

The tracking of all dependencies of a product, backwards chaining,  is done throughout the system in an un-compromised way, therefore an outstanding property is its 'extreme data lineage'. This is  amongst others used for evaluating the needs for re-processing and the quality control of data items and its dependencies including all input data and calibrations. The centralized role of the database is a key feature of Astro-WISE, which for large astronomical surveys, such as KiDS triggered notions like  "the universe as a spreadsheet" and "chaining to the Universe" \cite [(Valentijn \etal\ 2007)]{ADASSpaper2007}. The system is extensively described in a topical issue of Experimental Astronomy presenting 19 papers on its design, implementation and more advanced applications: \cite [Belikov \& Valentijn (2013)]{EAPapers2013} and \cite [Begeman \etal\ (2013)]{EAPaper2}.
   
The Astro-WISE approach of "living systems"  has been applied to several other domains  in  various projects of Target\footnote{http://www.rug.nl/target}, a public private collaboration of Artificial Intelligence, the Kapteyn Astronomical Institute, both the Computing and Medical Center of the University of Groningen, ASTRON, Target Holding, IBM, Oracle and many SME's.  
In this development the WISE approach, or its components, were used, but  always involving new data models. In fact, specific data federations are built for various domains, recognizing the specific level of standards, protocols and security for each of these domains.  Here, we briefly report on these  new data federations.

\section{Target}
Target is one of the largest public-private IT projects in the Netherlands. It specializes in the area of management and exploitation of large data volumes, commonly referred to as Big Data. Target has set up an expertise center at the University of Groningen where partners jointly conduct R\&D to develop innovative intelligent information systems that can efficiently process data and extract information from extremely large and structurally diverse scientific and other datasets.

A 10 Petabyte storage system and a supervising database has been installed in 2009 and has operated since then at the Center for Information Technology (\cite[Belikov \etal\  2011]{EAPaperTarget}). The functional requirements for this state-of-the-art ICT infrastructure were developed around the specific needs of the Target  projects, paying attention to scalability and cost-effectiveness. The data storage is governed by IBM GPFS, since this offered proven and effective scalability to Petabyte scales and beyond. Most projects made extensive use of an Oracle Real Application Cluster, which provides a system which is both powerful and reliable. 
GPFS proved to be optimizable for various access patterns ranging from many small files to less very large files. However, the optimization required the creation of specialized pools: a low latency disk pool with a capacity of 104 TB suitable for databases; a pool with large (4 TB) disks with a capacity of 2.500 TB suitable for bulk storage; a fast fiber-channel disk pool with a capacity of 104 TB  formatted for use by small files; a tape library with a capacity of 8.000 TB, for backup and near-online storage.

Amongst the lessons learned is the importance of the balance between reliability and performance in parallel file systems. The best performance is obtained by putting as many spindles as possible in one pool, but this comes at a loss of reliability due to a decrease in modularity. The Target infrastructure was optimized for performance, which has had disadvantages as projects moved into their production phases.

Processing for Target projects mainly used the 4.368 core Peregrine cluster.  The nodes are coupled using an 56 Gb/s Infiniband network. This network has both a high bandwidth and a low latency, especially suited for tasks using more than one machine.
\newpage
\section{KiDS}

\begin{figure}[h]  %[b]
%\vspace*{-3.3 cm}
%\vspace*{-3 cm}
\begin{center}
 \includegraphics[width=18.5 cm, natwidth=2713,natheight=1499]{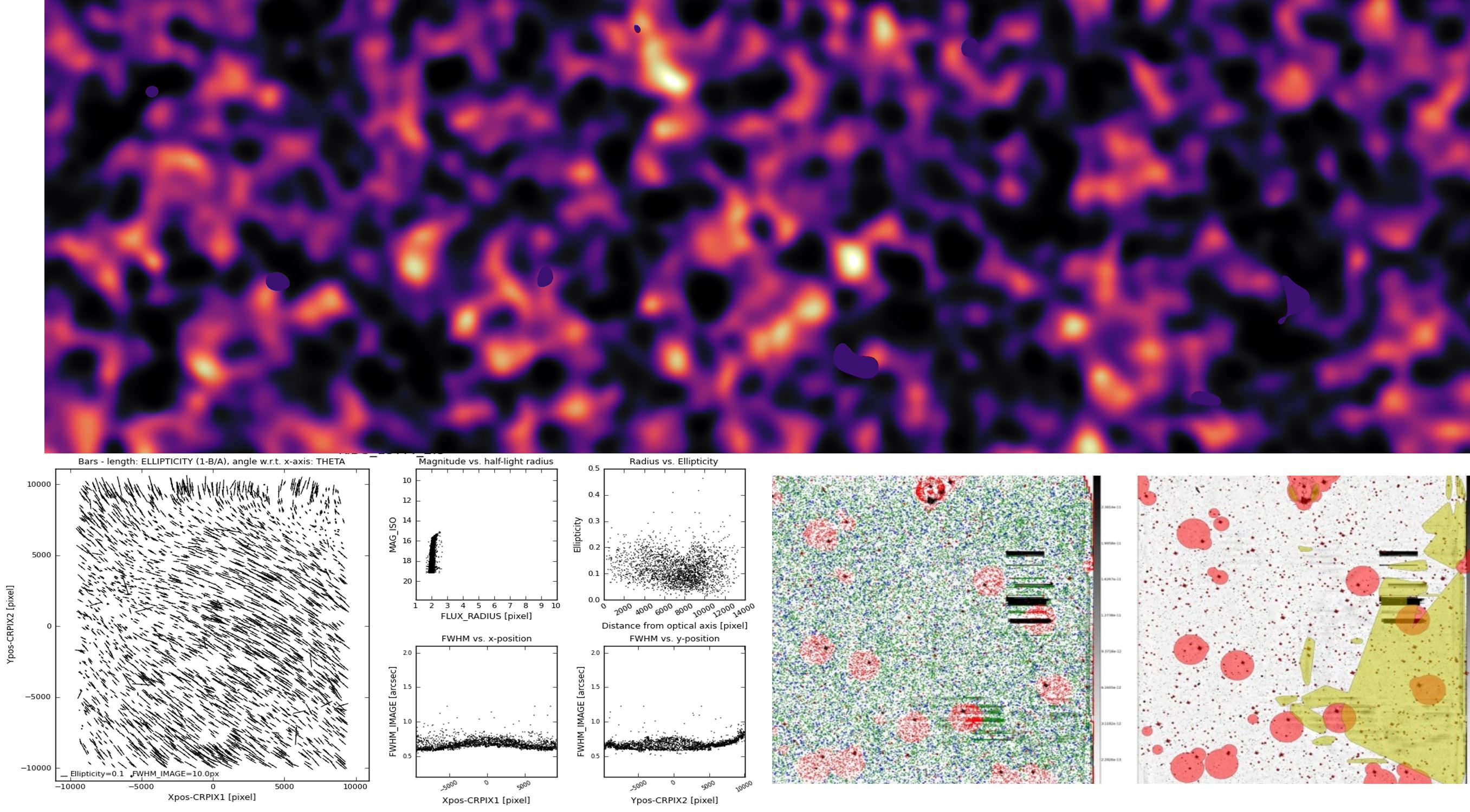} 
% \vspace*{-1.0 cm}
 \caption{KiDS dark matter map (ESO press release eso1641)
and quality control examples of image quality and the masking of bright stars.}
   \label{f:awe_qc}
\end{center}
\end{figure}
%\vspace{-2.0 cm}

The Kilo-Degree Survey (KiDS) is the largest ESO Public Survey with the OmegaCAM  256 Megapixel camera at the VLT Survey Telescope. KiDS covers 1500 square degrees in four bands.
Its aim is to study the large-scale distribution of dark matter and galaxies, using some of the sharpest wide-field image data available from ground-based telescopes.  Since 2011 KiDS is produced  with Astro-WISE in a data federation spanning the  Netherlands, Germany and Italy. The peer-to-peer architecture ensures production continues during local outages. Production monitoring and data sharing  are performed via queries to the WISE database. To date, the KiDS consortium has made 3 public data releases of 495 square degrees in collaboration with ESO and 6 internal consortium releases \cite[(de Jong \etal\ 2015)]{deJong2015}. This required quality assurance and control for tens of millions of stars and galaxies, tens of thousands of images and thousands of catalogs, amounting to 6 TB of end result data. The extreme data lineage facilitates the automatic quality assessment and diagnosis of failures. The assessment starts from end-products with easy back tracing to dependencies. In addition to automatic controls, a web interface  enabled a team of quality inspectors spread over Bonn, Naples, Groningen, Leiden and Edinburgh to perform efficiently visual inspection and manual quality annotations (Fig. 1). The Python binding of Astro-WISE allows astronomers to apply their code or work in a private "MyDB" partition.

\newpage
\section{MuseWISE}
\begin{figure}[h]  %[b]
%\vspace*{-2.5 cm}
\begin{center}
\hspace {2 cm}
 \includegraphics[width=19 cm, natwidth=1680,natheight=768]{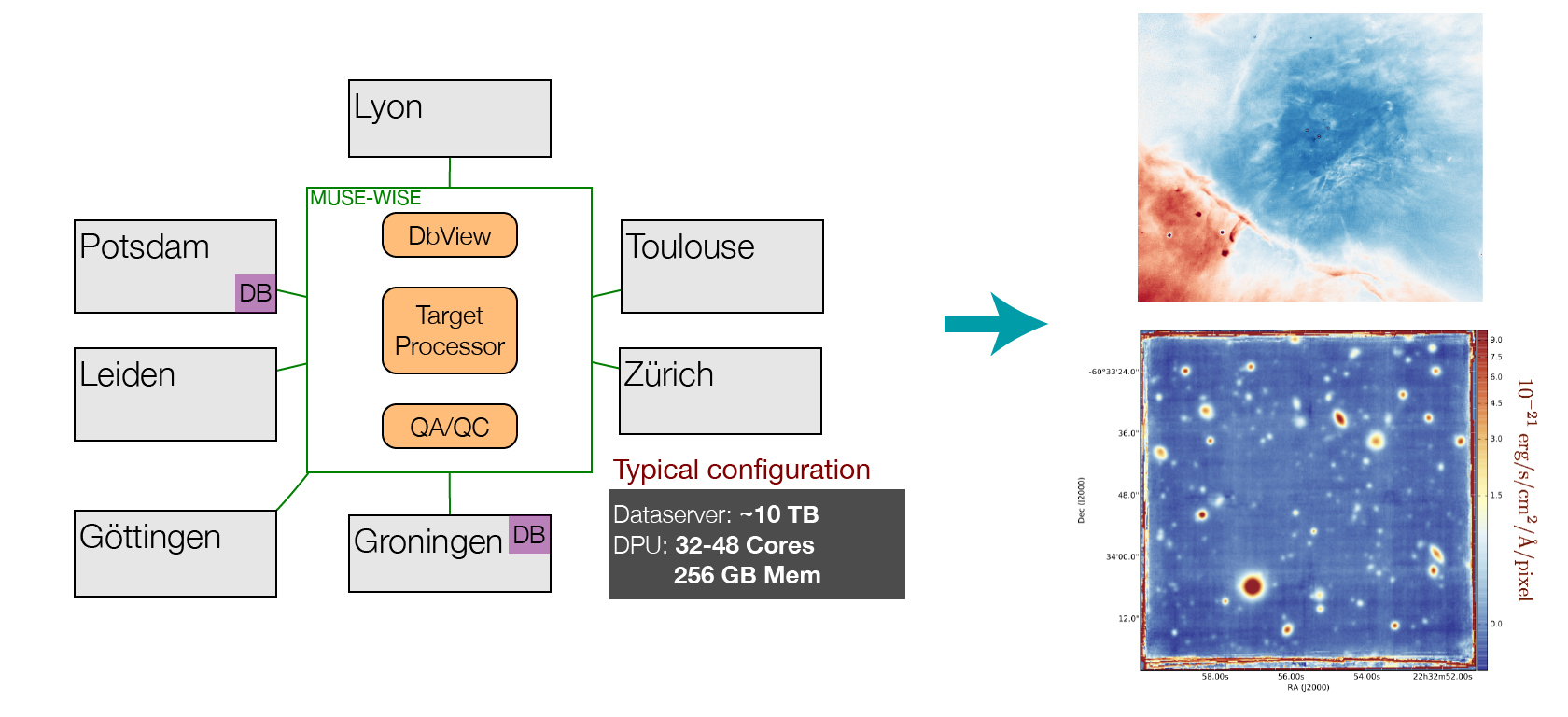}    
% \vspace*{-1.0 cm}
 \caption{Left: MuseWISE interlinks the 7 main nodes of the MUSE consortium. The DbView service allows access to data, and the results can be inspected through the quality control interface. The database is mirrored in Potsdam and Groningen to ensure minimal downtime and each node can reduce data independently. This infrastructure enables science of widely different types: from (top right) a panoramic view of the ratio [O III]/Hb of the Orion nebula (Weilbacher \etal\ 2015) to ultra-deep surveys of the distant Universe (bottom right), here shown by the white-light image of the Hubble Deep Field South  (Bacon \etal\ 2015).}
   \label{fig2}
\end{center}
\end{figure}

The Multi-Unit Spectroscopic Explorer (MUSE) is an integral-field spectrograph (IFS) developed for the VLT UT4. It saw first light in 2014 and with its 1'$\times$1' field of view sampled at 0.2'' it  offers an unprecedented view of the Universe.

In return for building the instrument, the MUSE team was awarded 250 nights of observing time on the VLT over a period of 5 years. This Guaranteed Time Observing (GTO) allocation offers a great opportunity to do break-through science but it also presents a data management challenge given the complex nature of the IFS data. To tackle this challenge, the MuseWISE system was developed as an extension of Astro-WISE to handle data from the MUSE GTO observations. This was integrated with a data quality assessment system developed in Toulouse in close collaboration with Groningen. 

The MuseWISE system has been operational since the commissioning of MUSE and now automatically ingests all GTO observations and provides automatic reductions of all data. As IFS data reduction is less well-developed than say, imaging, the production of final data-cubes is done outside MuseWISE but most MUSE GTO projects depend on MuseWISE for the timely execution of their data reduction. The infrastructure is also very well suited for the distribution of data, and for instance the $>100$GB datacube of the Orion nebula is published through MuseWISE with a sophisticated sub-cube service allowing access to relevant parts of the data.

\section{LOFAR Long Term Archive (LTA)}
LOFAR is presently the largest and most sensitive low-frequency 
radio telescope in the world. The array synthesis of the telescope requires impressive computational and 
storage facilities in order to process and archive the large volumes of data flowing through the system.
Together with ASTRON, CIT and OmegaCEN the Astro-WISE technology was used on the Target infrastructure to  create LOFAR's long-term archive (LTA - LoWISE), which  currently hosts  27 PB data. 
Presently, the LTA metadata respository, ingest servers, and search web-interfaces  
are hosted on the Target infrastructure  in Groningen. The  data storage is organized in a data federation with  SURFsara (Amsterdam), FZJ (J\"ulich, Germany) and PSNC (Poznan, Poland). As usual, the data federation is managed by the central database.  The central buffer can hold data for not more than a couple of weeks. Due to the strong pressure of off-loading the raw data from the observatory, (150 TB /week), the data federation proved to be essential to cope with outages and re-direct the dataflow over the three sites.

%\newpage
\section{GLIMPS}
\begin{figure}[h]  %[b]
%\vspace*{-3 cm}  %2.8
\hspace*{1 cm}
\begin{center}
 \includegraphics[width=18.5 cm, natwidth=1747,natheight=796]{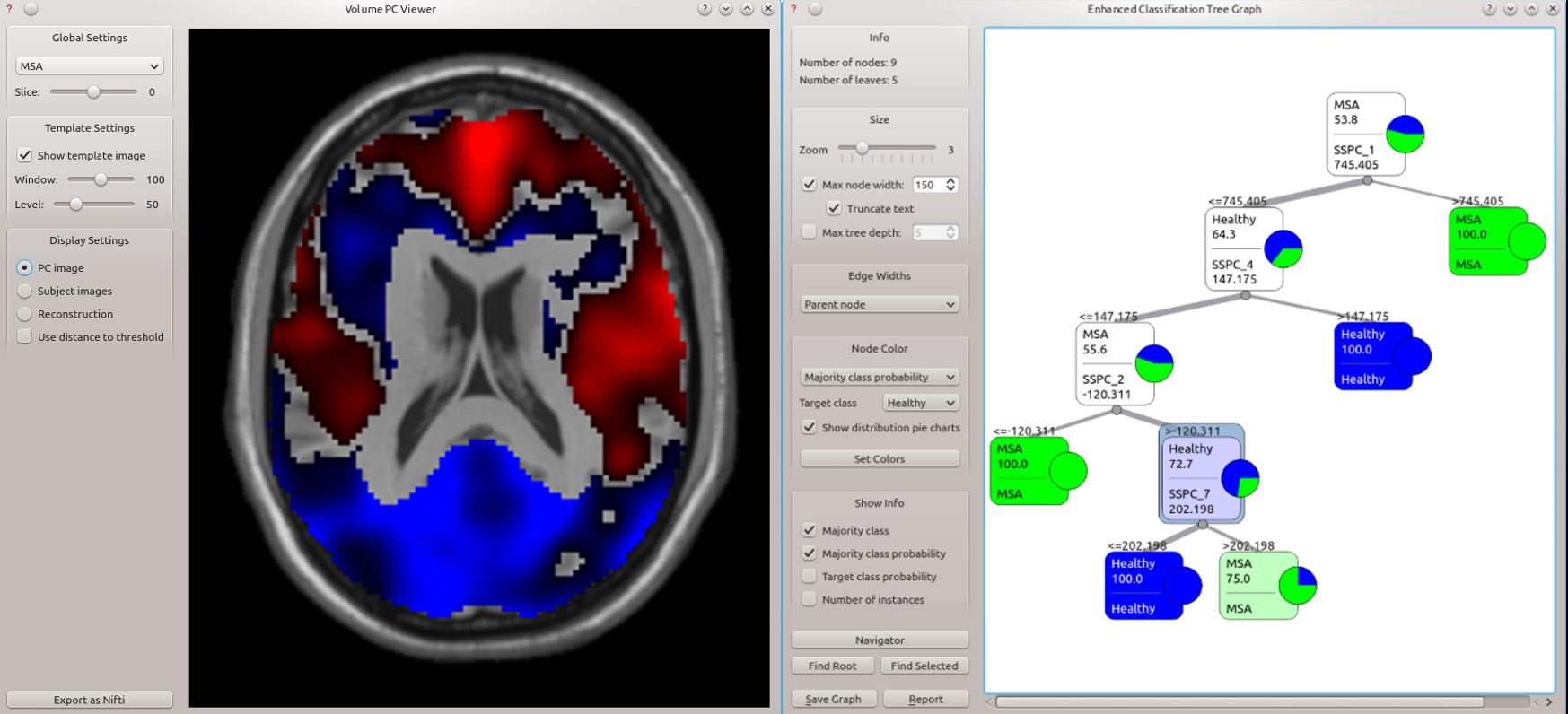}  % 19 cm
%\vspace*{-2.0 cm}
 \caption{Left: principal component thumbnail image corresponding to a selected tree node. Right: decision tree viewer (the lowest
   internal node is selected).}
   \label{fig-glipms}
\end{center}
\end{figure}
FDG-PET scans of the  human brain are ingested and distributed in a secure WISE environment. To aid clinicians in the diagnosis of various neurological diseases an image classification by decision trees has been developed. The construction of decision trees can be a
complex process, however, the resulting trees are typically simple and can be understood by users with little or no background in machine learning.  
%Decision trees have been applied to the problem of separating %
%healthy subjects from patients suffering from Parkinsonian
%syndromes, based on FDG-PET images. 
Principal component analysis is combined with the Scaled Subprofile Model  to obtain a set of subject scores for each subject which are used as features in decision
tree classification (\cite [Williams \etal\ 2016] {GLIMPSpaper}). 
A problem is that the subject scores are not very intuitive. 
%so it is difficult to extract meaningful information directly 
%from the resulting tree.
To amend this problem, the tree visualization can be augmented by
internal nodes with thumbnails of the corresponding principal
component images.  An example is shown in Fig.~ 3.
The system has been implemented as an interactive viewer via an
extension of the Orange data analysis software. The additional
visualizations can improve a users understanding of the
classification process, and provide insight into the knowledge that has been captured by a decision tree.

\section{MONK}

\begin{figure}[h]  %[b]
%\vspace*{-2.0 cm}
\begin{center}
%\hspace {3 cm}
 \includegraphics[width=10 cm, natwidth=1254,natheight=880]{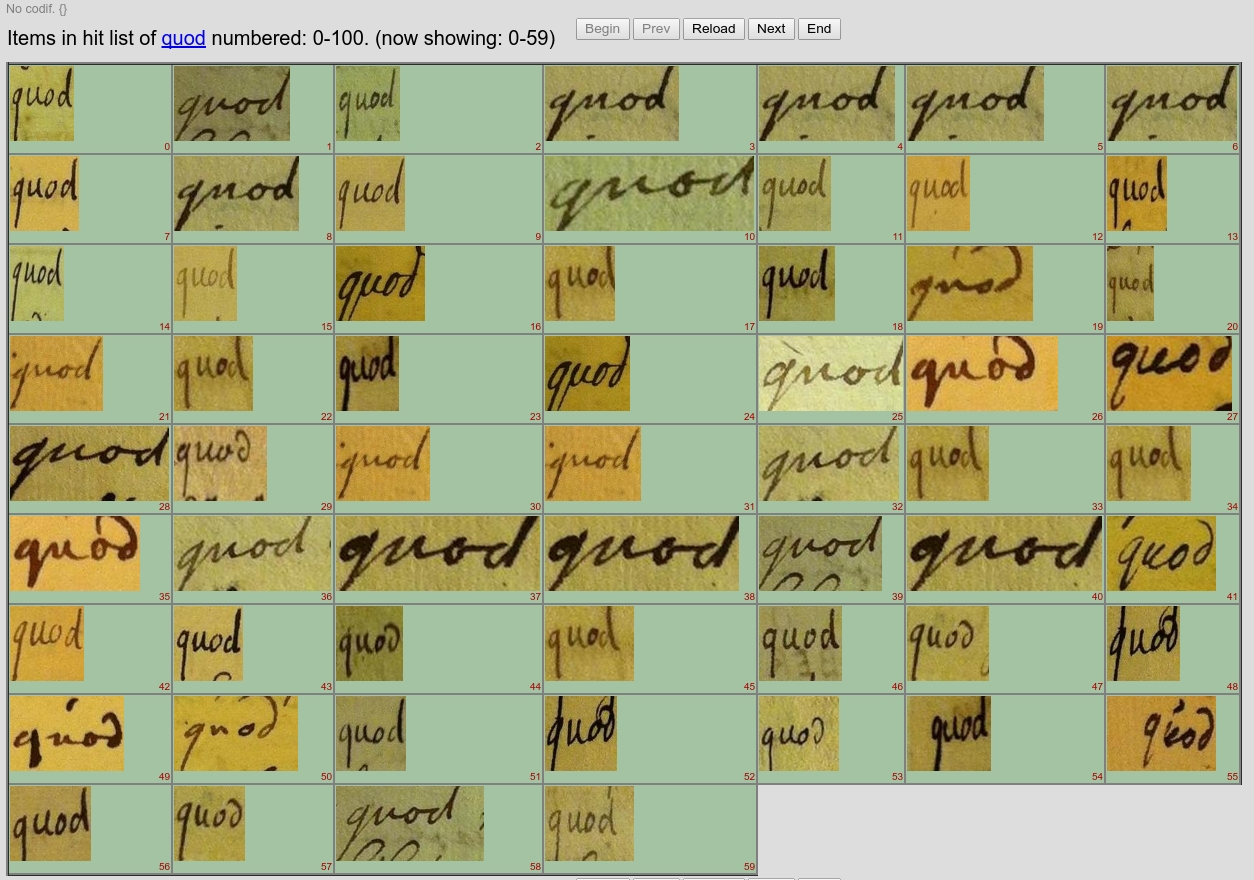} 
% \vspace*{-1.0 cm}
 \caption{MONK hit list for the Latin word 'quod' from a 17th century collection of scholarly letters by different authors.
Monk contains 400 documents, 76k pages, 1.9M lines of text, 777k harvested labels and 127k lexical words of collections from: the Dutch National Archive; Czech National Archive; Dutch Royal Library; City Archives of Louvain, Belgium; Harvard University Yenchin (Chinese) collection; the Dead Sea Scrolls and many others.}
   \label{fig1}
\end{center}
%\vspace {-0.5 cm}
\end{figure}

The goal of the Monk project is to enable direct access to scanned images of handwritten historical collections, by searching for keywords depicted anywhere in these images. Since standard 'OCR' (optical character recognition) is not possible on ancient text images, new methods needed to be developed.
The problem entails a multitude of historical script styles and ancient languages, including severe image problems such as ink fading, bleeding, curvilinear text lines and complex background textures. Therefore, a type of machine learning is required that allows starting from scratch, with zero knowledge, learning from gradual experience. This was realized by using interactive labeling and machine learning using high-performance computing. An architecture was developed that exposed individual word-image candidates over the internet in a way which is convenient to both programmers and scholars. In order to achieve this, the extreme position was taken that all disk I/O should be done on a normal, familiar (i.e., Posix) file system. This was a risky experiment, but we have realized a storage architecture that is able to handle up to two billion files. Currently, a collection of 1.3 billion files is realized, the majority consisting of character and word candidate images (\cite [van Oosten en Schomaker 2016]{MONKpaper}). The project yielded new concepts in machine learning, notably the insight that a prolonged iteration over the processing sequence N x 'recognition, then reranking', yields an effective retrieval engine. Learning appears to occur in phase transitions, such that the number of harvested word labels suddenly steeply increases at the point where a critical number of examples is reached - 
% We have dubbed this iterative approach
the 'Fahrkunst principle', improving both the discrimination of classes and the ranking of hit-list instances.
%The necessity of the principle comes from the dual need of 
% optimizing both the separability of classes and the 
% prototypicality of hit-list instances. 
%
\section{Health data infrastructure: LifeLines, Genome of the Netherlands and BBMRI-NL}
The Target infrastructure was also deployed for  human subject research; in particular Lifelines, the largest biobank of the Netherlands(165,000 individuals- http://lifelines.net), Genome of the Netherlands, the first  national whole genome project (http://nlgenome.nl) and the Dutch national node of Europe's Biobanking and Biomolecular Research Infrastructure (BBMRI). Target has been essential because of the rapid upscaling in large biobanks (collections of samples and measurements of human subjects) and new high throughput measurement techniques, e.g. allowing the measurement of all 3.1 billion human DNA elements in one experiment. Life scientists suddenly required large data processing capability that was not available elsewhere at that time, e.g. 'nlgenome' contains 500 TB of data. 

Special to human subject research are the ethical/legal issues surrounding privacy of the individuals donating the data/samples. This required a tightly controlled data federation including dedicated secure 'workspaces' where a consortium of researchers can collaborate on the data without concerns over unauthorized data access. This has also lead to new methods of data federation where data catalogues can be publicly searched, but imposing ethical approval before data analysis can commence. This is supported via new software implementations in the MOLGENIS platform (\cite[Swertz \etal 2010]{}) which is now used for the European BBMRI data and sample Directory (515 biobanks, 60 million samples) (\cite[Holub \etal\  2016]{}). This project has resulted in $>40$ new publications and $>10$ prestigious personal grants, including top publications in sequencing of the gut microbiome 
(Zhernakova \etal\  2016), gene expression and DNA methylation. 

% 
% page break here
%
%\newpage
\section{Target- societal impact }
The agenda of Target included an extensive economic stimulation programme, triggering new jobs, investments and spin-off companies. Elkoog BV built and launched its new product Crowdynews, which finds, selects, ranks, filters, and publishes high quality content from an ocean of tens of millions of produced items every day (e.g. Twitter) in every language of the world for hundreds of renowned online news publishers around the globe. Today, Crowdynews serves over ten million unique readers on a daily basis and has 60 employees (55 new jobs). \\
Target Holding has created over 12 spin-off companies, ranging from internet mining engines (Songa, Dataprovider), to user platforms in health (e-vitality). With a renowned international publisher, Target Holding is developing a machine learning approach for selecting the referees for the 0.5 million papers annually submitted to their journals.\\
To date, Target has created over 150 new jobs, and triggered over 24 million Euro of new investments, including a state of the art digital planetarium theatre downtown Groningen. It is still a dream to connect the international planetarium community in a data federation, but the Open Space (openspaceproject.com) initiative might be a first step. 

\newpage
\section{Euclid}
\strut
\vspace{-1 cm}
\begin{figure}[h]  %[b]
%\vspace*{-3.5 cm}
\begin{center}
\hspace* {2 cm}
\includegraphics[width=14 cm, natwidth=1015,natheight=781]{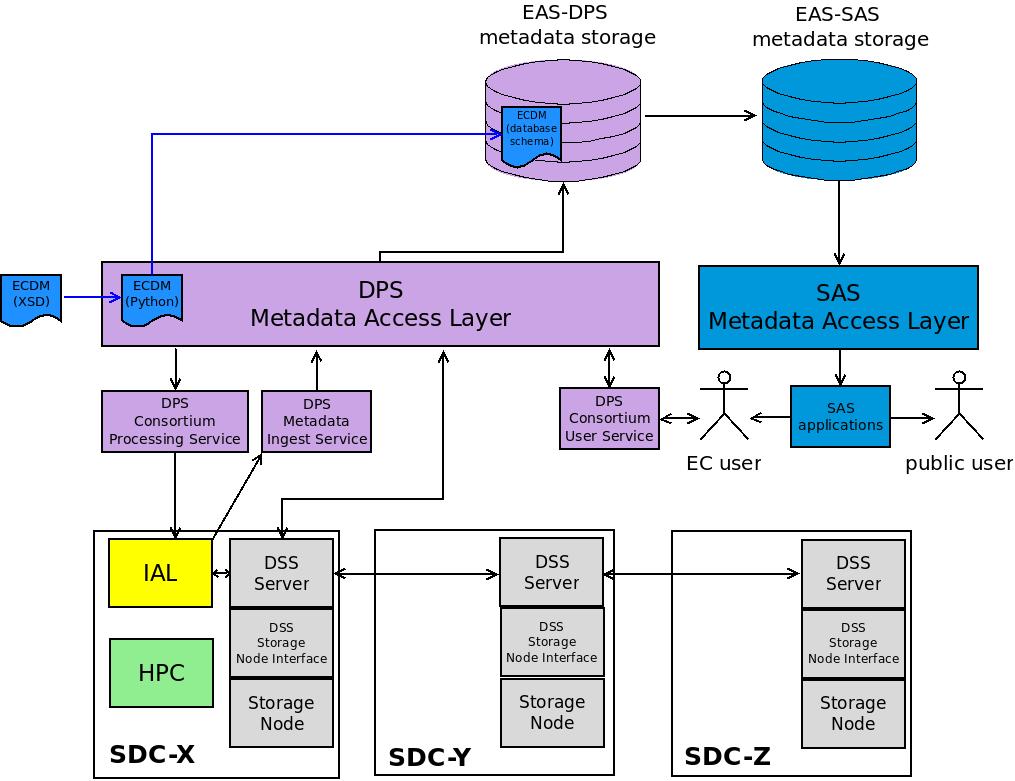} 
 \caption{Lay-out of the datahandling of Euclid, with a glue role for the EAS}
   \label{fig5}
\end{center}
\end{figure}

The Euclid Archive System (EAS) forms the core of the data processing system (\cite [Nieto \etal\ 2017]{Euclidpaper}) for this milestone ESA space mission. Apart from the total data volume,  which will reach well over 20 PB, one of the main Euclid challenges is the very high accuracy required for the results, which will involve elaborate quality control and numerous reprocessing of the data. To achieve this, the data-centric and data lineage  approach of Astro-WISE will be deployed, allowing the reduction of unnecessary reprocessing and providing users with quality assessment of  data products throughout the mission and throughout the different steps in the data processing. 
The EAS will store data files in a distributed grid of data storage servers and metadata in a centralized database. The data model of Euclid will be complex due to the large number of processing steps and the detailed modules involved. The EAS has to provide uninterrupted metadata flow from all processing steps, distributed over 9 National Science Data Centers (SDCs), binding all data products. Dedicated metadata browsing services and quality services will allow pipeline developers and calibration scientists to refine processing and trace any effects appearing in the data processing chain. 
The final step in the data processing is a public data release via the Science Archive System, hosted by ESA. 
The EAS is a joint development of the Euclid Consortium represented by the Dutch SDC  and  
ESAs SDC. 

%\newpage

\section {Data Federations}
One of the most important lessons learned from the Target initiative is that modern distributed big data projects require a strong attention to project management, data modelling, the various  levels of standards and protocols in different domains, the wide range of requirements on security and privacy, the zoo of hardware at different datacenters, the various views on security and operational constraints of data centers, and finally and most particularly the appreciation of the very different sociology in different communities. A single "heal the world type of system"  will not work. Instead a common approach, with common libraries and concepts will work, but only when recognizing these differences by installing domain specific data federations. In Target this proved to be the most efficient approach.\\
It is demanding, but rewarding, to initiate extensive data modeling at the start of a project. In Astro-WISE this is approached inherently in the definition of its Python classes, but other approaches, such as the XML modeling as done in MUSE or the XSD classes delivering XML objects (Euclid)  are also satisfactory solutions. For experimental projects like LOFAR, where  additional calibration techniques are  being developed during operations this is clearly a big hurdle, and it will take more time to install the data modeling for the eventually settled data processing. For big projects like Euclid,
% involving 8 datacenters spread over Europe,
 the variety of hardware and views on security by datacenters implied the development of a large set of common access layers, in order to abstract this away.

\strut \\
\textit{Target is supported by Samenwerkingsverband Noord Nederland, European Fund for Regional development, Ministery of economic affairs, Provinces of Groningen and Drenthe.}

\end{document}